\documentclass[]{interact}

\usepackage{epstopdf}
\usepackage[caption=false]{subfig}

\usepackage[numbers,sort&compress,merge]{natbib}
\bibpunct[, ]{[}{]}{,}{n}{,}{,}
\renewcommand\bibfont{\fontsize{10}{12}\selectfont}

\theoremstyle{plain}

\theoremstyle{definition}

\theoremstyle{remark}

\newcommand{\redlaser}{\ensuremath{f_{729}}}
\newcommand{\redbeat}{\ensuremath{f_{729}^{\text{Beat}}}}
\newcommand{\redn}{\ensuremath{n_{729}}}

\newcommand{\uslaser}{\ensuremath{f_{\text{US}}}}
\newcommand{\usbeat}{\ensuremath{f_{\text{US}}^{\text{Beat}}}}
\newcommand{\usn}{\ensuremath{n_{\text{US}}}}

\newcommand{\qcllaser}{\ensuremath{f_{\text{QCL}}}}

\newcommand{\sfglaser}{\ensuremath{f_{\text{SFG}}}}
\newcommand{\sfgbeat}{\ensuremath{f_{\text{SFG}}^{\text{Beat}}}}
\newcommand{\sfgn}{\ensuremath{n_{\text{SFG}}}}

\newcommand{\rep}{\ensuremath{f_{\text{Rep}}}}
\newcommand{\ceo}{\ensuremath{f_{\text{CEO}}}}

\begin{document}


\title{Frequency stabilisation and SI tracing of mid-infrared quantum-cascade lasers for precision molecular spectroscopy \footnote{Dedicated to Prof. Wim Ubachs on the occasion of his 65$^\text{th}$ birthday.}}

\author{
\name{Mudit Sinhal\textsuperscript{a}, Anatoly Johnson\textsuperscript{a} and Stefan Willitsch\textsuperscript{a}\thanks{Corresponding author: Stefan Willitsch. Email: stefan.willitsch@unibas.ch}}
\affil{\textsuperscript{a}Department of Chemistry, University of Basel, Klingelbergstrasse 80, 4056 Basel, Switzerland}
}

\maketitle

\begin{abstract}

The advancement of technologies for the precise interrogation of molecules offers exciting possibilities for new studies in the realms of precision spectroscopy and quantum technologies. Experiments in these domains often address molecular vibrations in the mid-infrared (MIR) spectral region, thus generating the need for spectrally pure and accurate MIR laser sources. Quantum cascade lasers (QCLs) have emerged as flexible sources of coherent radiation available over a wide range of MIR frequencies. Here, we demonstrate a robust approach for the simultaneous linewidth narrowing, frequency stabilisation and absolute frequency referencing of MIR QCLs all of which are prerequisites for precise spectroscopic experiments. Following upconversion of its radiation to the visible domain, we implement a phase lock of the QCL to a linewidth-narrowed optical frequency comb which is referenced to a remote SI-traceable primary frequency standard via a fibre link for absolute frequency calibration. To achieve a reliable frequency counting of the beat note between the QCL and the OFC, we employ redundant tracking oscillators and demonstrate a frequency instability of $5\times10^{-13}$ at 1~s and $2\times10^{-14}$ at 1000~s integration time, limited by the accuracy of our remote reference. 

\end{abstract}


\section{Introduction}

The recent advancements in techniques for the precise spectroscopic investigation of molecules has enabled new applications such as improved tests of molecular theory \cite{patra20a, holsch19a}, precise determinations of molecular energy levels and properties \cite{semeria20a, melosso21a}, the development of new frequency standards and molecular qubits \cite{najafian20a, karr14a}, accurate determinations of fundamental constants and their putative variations \cite{moretti13a, jansen14a, hudson06a, shelkovnikov08a, schiller05a, kajita14a, kortunov21b, patra20a} and investigations of possible physics beyond the standard model \cite{salumbides13a} as demonstrated in recent works by Ubachs and co-workers \cite{patra20a, holsch19a,melosso21a,jansen14a,salumbides13a, germann21a,diouf22a,melosso21a, diouf21a,lai21a, lai21b,hussels22a,cozijn22b,ryzner22a, malicka21a}.


Precise spectroscopic studies of rovibrational transitions in molecules in the mid-infrared (MIR) frequency domain necessitate the development of accurate, precise, and spectrally narrow MIR lasers. Quantum cascade lasers (QCLs) have proven to be attractive MIR sources capable of operating close to room temperature and emitting intense coherent tunable radiation. Their availability over a wide range of frequencies from 3--25~$\mu$m has also enabled their use in trace gas sensing \cite{weidmann04a, hundt18a}, in free-space MIR optical communications \cite{liu15b, martini05a}, and in analytical applications \cite{deutsch14a}. 
Free running QCLs can exhibit linewidths of few tens of kilohertz to many megahertz \cite{schilt13a} which is often insufficient for state-of-the-art high-resolution experiments. Therefore, suitable schemes for linewidth narrowing need to be implemented, ideally combined with an absolute frequency stabilisation as is often required in precision measurements. In the past, reduction of QCL linewidths by locking to high-finesse cavities \cite{fasci14a, taubman02a} and by phase-locking to MIR references \cite{sow14a} have been demonstrated. In addition, locking and referencing schemes based on nonlinear processes have been widely employed. These include injection locking \cite{borri12a} and phase locking \cite{galli13a} of QCLs to difference frequencies derived from near-infrared (NIR) lasers as well as stabilisation by locking upconverted QCL frequencies to ultra-low expansion (ULE) cavities \cite{hansen15a} and to optical frequency combs (OFCs) \cite{hansen13a, argence15a, mills12a}.   

In the present work, we demonstrate an alternative stabilisation scheme of a QCL by an OFC the linewidth of which is narrowed by a tight lock to an ultralow-expansion (ULE)-cavity-stabilised master laser at 729~nm. The QCL operates around a wavelength of 4.5~$\mu$m tailored for studying dipole-forbidden infrared excitations in the N$_2^+$ molecular ion \cite{germann14a, najafian20b}. In addition, the OFC is referenced to an SI-traceable ultrastable (US) laser disciplined to the Swiss primary frequency standard at a remote location and disseminated to our laboratory via a phase-stabilised fibre network \cite{husmann21a}. Since the OFC does not have emission in the MIR spectral region, we employ a sum-frequency generation (SFG) process in order to upconvert the QCL output to visible domain which is then used to phase-lock the QCL to the OFC. Due to the low powers of both the OFC output and the radiation produced by SFG, the generation of beat notes with an adequate signal-to-noise ratio for direct and reliable frequency counting poses challenges. We thus implement redundant tracking oscillators \cite{johnson15a, udem98a} which are phase-locked to the beat note with different locking bandwidths. This measure results in a better filtering of the signal, enabling a real-time cycle-slip-free counting of the beat note and thus determination of the QCL frequency.  

The present paper is organised as follows. The core elements of the present approach for stabilising the OFC and QCL are presented in sec. \ref{sec:stab_ofc} and \ref{sec:stab_qcl}, respectively. The performance of the overall scheme is evaluated in sec. \ref{sec:perf} and discussed in sec. \ref{sec:disc}. The paper finishes with an outlook and conclusions in sec. \ref{sec:conc}.

\section{Frequency stabilisation of the optical frequency comb}
\label{sec:stab_ofc}

\begin{figure}[ht!] 
    \centering
    \includegraphics[height=0.9\linewidth,trim={0cm 0cm 0cm 0cm},clip]{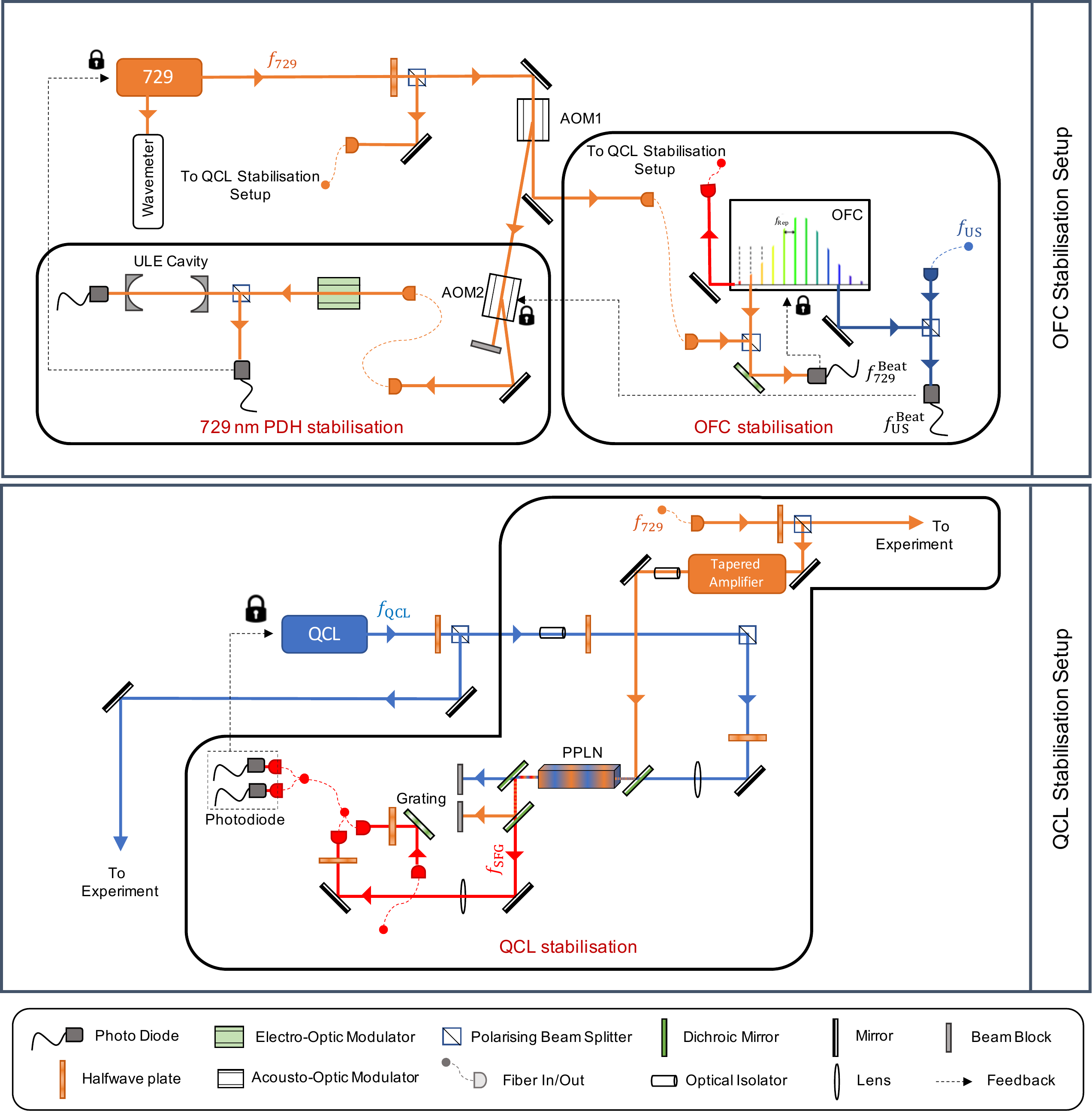}
    \caption[Stabilisation Setup]{Stabilisation setups for the optical frequency comb (OFC) and the quantum cascade laser (QCL). The OFC was phase locked to an optical-cavity-stabilised 729~nm master laser with emission frequency \redlaser{}. An ultrastable (US) reference laser at a frequency \uslaser{} corresponding to 1572.06~nm, generated at the Swiss Institute of Metrology METAS, was transferred to our laboratory at the University of Basel via a 123~km long phase-stabilised fibre link, where it was employed to measure and compensate drifts of the ultralow-expansion (ULE) cavity for the master laser. The beat notes of the OFC with the 729~nm master laser and the US laser (\redbeat{} and \usbeat{}, respectively) were detected using beat-detection units. The mid-infrared (MIR) QCL, emitting at a frequency \qcllaser{}, was then phase locked to the OFC after upconversion of its frequency to the visible domain (\sfglaser{}) by sum-frequency generation (SFG) with amplified emission from the master laser in a periodically poled lithium niobate (PPLN) crystal. The beat note between the QCL and the OFC was generated in a balanced photo-detection setup. See text for further details.}
    \label{fig:Stab_Setup}
\end{figure}

A schematic of the complete stabilisation setup is presented in fig. \ref{fig:Stab_Setup}. In a first step, a 729~nm external-cavity diode laser (ECDL, Toptica DLPro) was locked to an ULE cavity with a finesse of $280'000$ and a linewidth of 5.3~kHz using a Pound-Drever-Hall (PDH)  scheme (``729~nm PDH stabilisation'' in fig. \ref{fig:Stab_Setup}) \cite{black01a}. The setup for the 729~nm stabilisation consisted of a first acousto-optic modulator (AOM) (labelled ``AOM1'') operating at $\sim 1.1$~GHz employed to bridge the frequency gap between the laser frequency, \redlaser{}, and the transmission modes of of the ULE cavity spaced 1.5~GHz apart. A second AOM (labelled ``AOM2''), with a centre frequency of 200~MHz, allowed for a fine control over the laser frequency.    

\begin{figure}[ht] 
    \centering
    \includegraphics[width=0.9\linewidth,trim={0.5cm 4.cm 0.5cm 4.5cm},clip]{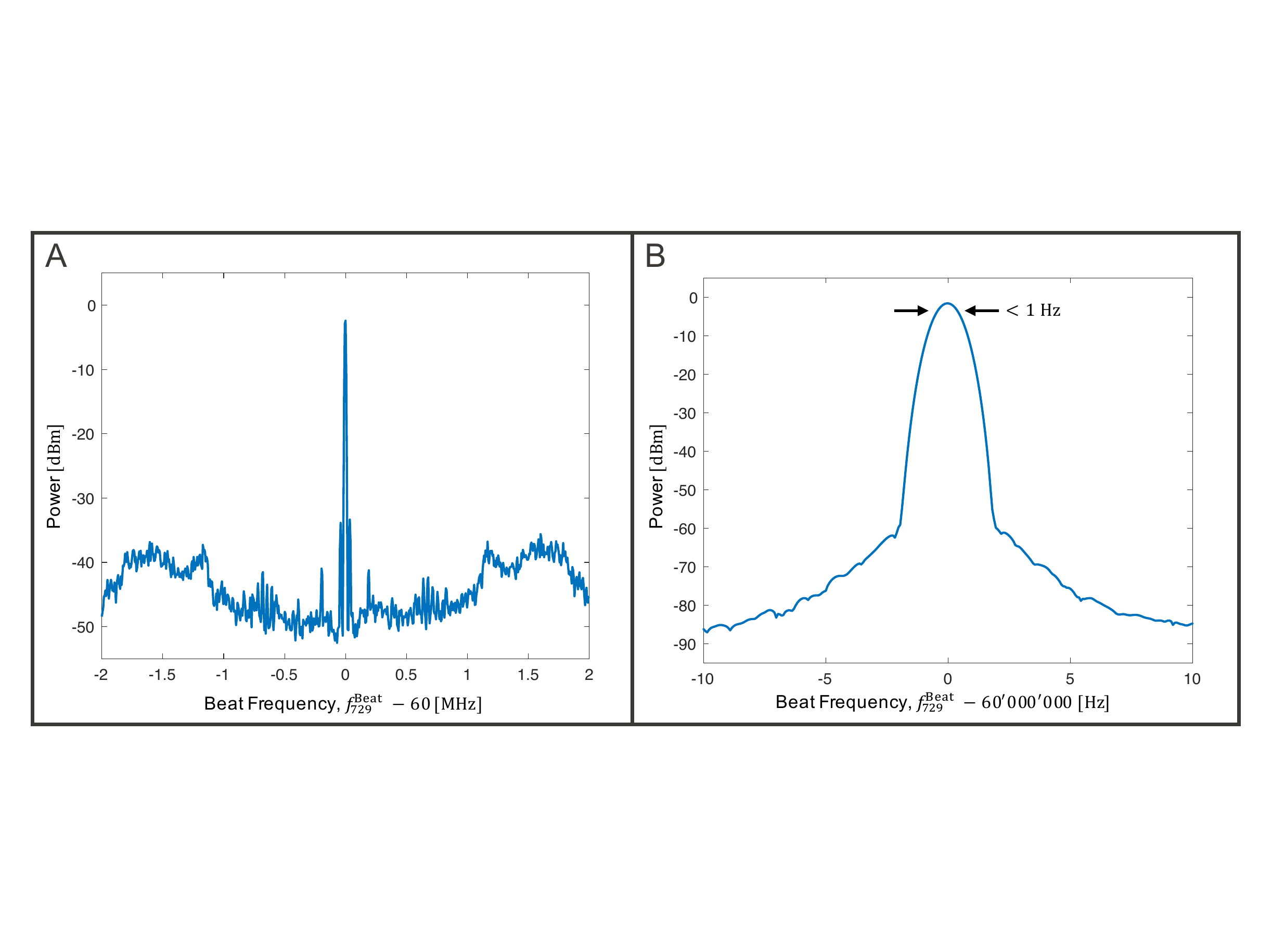}
    \caption[Locking Results]{A) In-loop beat note of the 729~nm master laser with the OFC. The beat note exhibits a signal-to-noise ratio $>35$~dB measured with a 10~kHz resolution bandwidth. B) Magnification of the central feature in panel A. The linewidth of the beat note of the locked laser was determined to be $<1$~Hz limited by the resolution bandwidth of the spectrum analyser employed.}
    \label{fig:729lock}
\end{figure}

In a second step, part of the frequency-stabilised light at 729~nm (the $0^{\text{th}}$ order of  ``AOM1'') was used to phase lock and narrow a femtosecond OFC (Menlo Systems FC1500-ULN) (``OFC stabilisation'' in fig. \ref{fig:Stab_Setup}). For this purpose, a fully-fibre-coupled beat-detection unit (Menlo Systems BDU-FF) was employed in order to generate a beat note, \redbeat{}, between the 729~nm laser and emission from the OFC centred at centred at the same wavelength which was generated by amplifying, frequency shifting and frequency doubling a portion of the OFC radiation (Menlo Systems M-VIS extension). A phase-locked loop was then engaged in order to lock the repetition rate of the OFC, \rep{}, at a frequency of $\sim250$~MHz. The tight lock of the OFC to the narrow-linewidth master laser implied a narrowing of the linewidth of the individual comb teeth. The carrier-envelope-offset (CEO) frequency, \ceo{}, of the OFC was detected using $f-2f$ interferometry and phase locked to a Global Positioning System disciplined (GPSD) Rb clock (SRS FS725 Rb standard disciplined to Symmetricom GPS-500), which also served as the local oscillator. Fig. \ref{fig:729lock}A shows the spectrum of the in-loop beat between the 729~nm laser and the OFC when the OFC is phase-locked to the 729~nm laser. As shown in fig. \ref{fig:729lock}B, the beat note with frequency \redbeat{} exhibits a linewidth of $\sim1$~Hz (full width at half maximum, FWHM) limited by the 1~Hz resolution bandwidth of the  spectrum analyser employed (Teledyne T3SA3200) and indicating that the frequency of the relevant comb tooth tightly follows the master laser.

The \rep{}, \ceo{} and \redbeat{} frequencies were counted using dead-time-free counters (Menlo Systems FXM50) referenced to the GPSD Rb clock. From these measurements, the frequency of the 729~nm laser could be determined as, 
\begin{equation} \label{eqn:729GPS}
    \redlaser{} = 2\ceo{} + \redn{}\rep{} + \redbeat{},
\end{equation}
where, $\redn{} = 1'644'151$, was the mode number of the OFC used to generate the beat with the 729~nm laser and the factor of 2 with \ceo{} appears because the spectrum of the OFC was doubled for the beat-note generation. 

Ageing of the ULE glass \cite{dube09a} causes a nonlinear drift of the cavity length resulting in a drift of the frequency of the 729~nm laser and consequently a drift of the OFC repetition rate. This resulted in a poor long-term frequency stability of the lasers. In order to measure and compensate these drifts, the OFC was additionally referenced to an US NIR reference laser operating in the telecom L-band at 1572.06~nm. The US reference was generated at the Swiss Federal Institute of Metrology (METAS) in Berne and disseminated to our laboratory at the University of Basel via a phase-stabilised optical-fibre link \cite{husmann21a}. The frequency of the US laser, \uslaser{}, was accurately determined at METAS by a comparison to an active maser contributing to the realisation of the Swiss Coordinated Universal Time, UTC(CH), resulting in its SI traceability. Further details about the US laser generation and its dissemination can be found in ref. \cite{husmann21a}. 

\begin{figure}[ht] 
    \centering
    \includegraphics[width=0.9\linewidth,trim={0.5cm 4.2cm 0.5cm 4.2cm},clip]{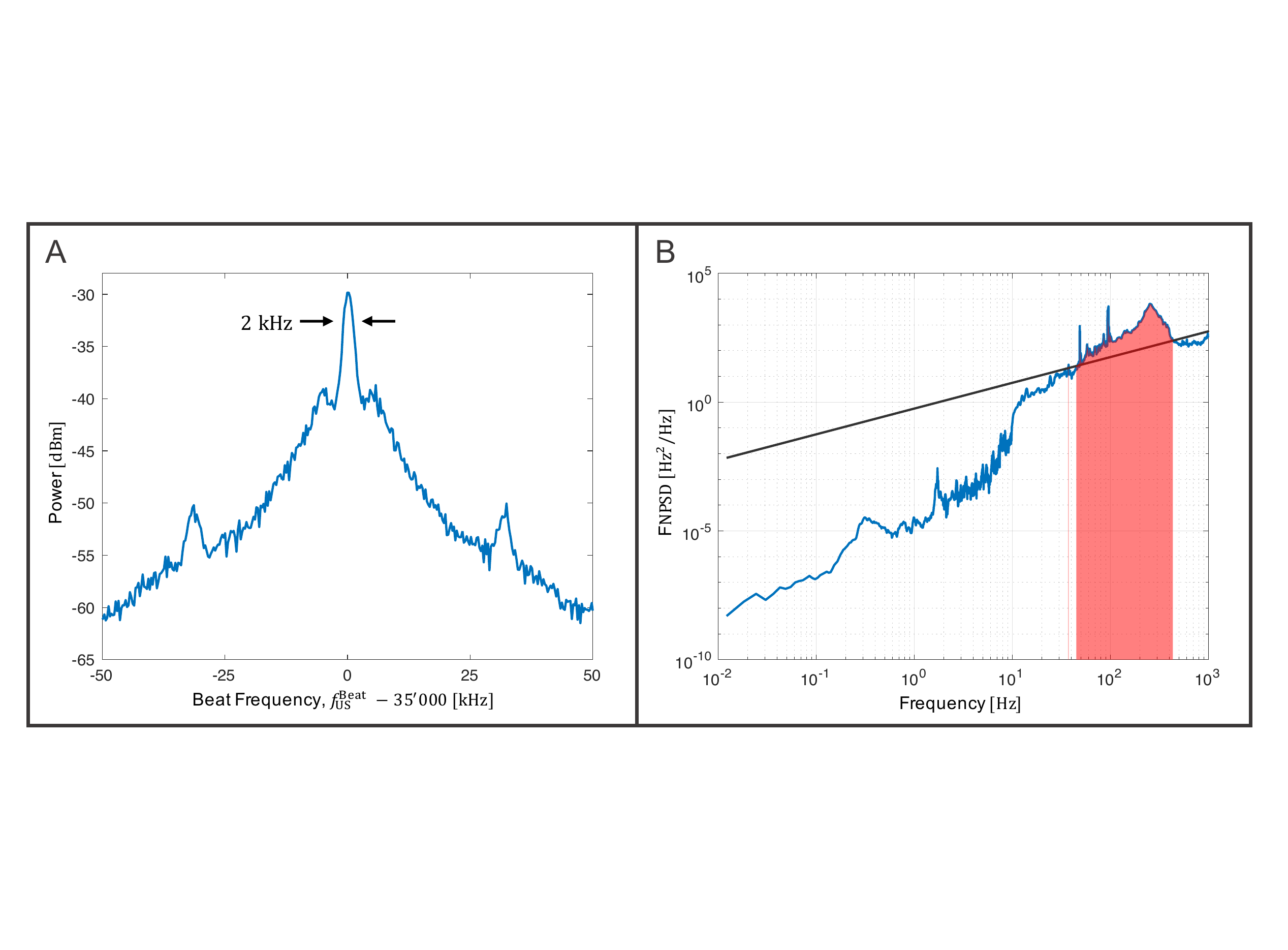}
    \caption[Locking Results]{A) Out-of-loop beat note between the stabilised OFC and the US laser generated at METAS delivered to the University of Basel via 123~km of optical fibres. The beat note exhibits a linewidth of 2~kHz. B) Frequency-noise power spectral density (FNPSD) of the round trip phase-noise-cancellation (PNC) system measured in Berne. The portion of the spectrum above the beta-separation line (red shaded area) contributes to the noise of the linewidth observed in panel A.}
    \label{fig:USlock}
\end{figure}

Fig. \ref{fig:USlock}A shows the beat note with frequency \usbeat{} between the stabilised OFC and the US laser. A FWHM linewidth of 2~kHz could be observed, dominated by the noise of the 123~km long fibre segment used to deliver the US laser from Berne to Basel \cite{husmann21a}. Fig. \ref{fig:USlock}B shows a measurement of the frequency-noise power-spectral density (FNPSD) of the round-trip phase-noise cancellation (PNC) system employed in order to compensate the fibre noise. Here, the method of the beta-separation line (black trace) was used to determine the linewidth of the radiation \cite{di10a}. The thus determined FWHM linewidth of the US laser at Basel was estimated to be $\sim1.85$~kHz from the area of the slow-modulation region (red-shaded region in Fig. \ref{fig:USlock}B)  \cite{di10a} indicating that the observed linewidth of \usbeat{} was due to the fibre noise and that the linewidths of the OFC and the US lasers could be expected to be substantially smaller than 2~kHz. 

The beat note with frequency \usbeat{} between the OFC and the US laser was counted on another channel of the  frequency counter referenced to the GPSD-Rb clock. A slow feedback was subsequently applied on ``AOM2'' in order to compensate the drifts of the ULE cavity. From the counter readings, the frequency of the US laser could be expressed as, 
\begin{equation}  \label{eqn:fus}
    \uslaser{} = \ceo{} + \usn{}\rep{} + \usbeat{},
\end{equation}
where, $\usn{} = 762'808$, is the index of the comb tooth beating with the US laser. 

The precision of the measurement of the frequency \redlaser{} of the 729~nm master laser using eq. (\ref{eqn:729GPS}) was limited by the instability of the GPSD Rb clock amplified by the large multiplication factor \redn{}. However, using eq. (\ref{eqn:fus}) the frequency of the 729~nm laser could be determined both accurately and precisely by eliminating \rep{} and expressing \redlaser{} in-terms of the US laser frequency as \cite{telle02a},
\begin{equation} \label{eqn:729US}
    \redlaser{} = (2 - x)\ceo{} + x(\uslaser{} - \usbeat{}) + \redbeat{},
\end{equation}
where, $x = \redn{}/\usn{} \approx 2.15$, is the ratio between the index of the comb teeth for the beats with the 729~nm and the US laser, respectively. 

\section{Frequency stabilisation of the quantum-cascade laser}
\label{sec:stab_qcl}

The bottom panel of fig. \ref{fig:Stab_Setup} shows the setup for locking the QCL to the frequency-stabilised OFC. The QCL used in this work was a continuous wave (CW) distributed-feedback (DFB) device (Adtech Photonics) operating at a temperature of $16^\circ$C and emitting radiation with frequency $\qcllaser{}$ corresponding to a wavelength of $\approx 4.57~\mu$m with an output power of 400~mW. For SFG, about 100~mW of the QCL radiation were overlapped with amplified 729~nm light from the master laser in a periodically-poled lithium-niobate (PPLN) crystal resulting in the generation of 50~$\mu$W of sum-frequency radiation at 629~nm with frequency \sfglaser{} given by, 
\begin{equation} \label{eqn:sfg}
    \sfglaser{} = \qcllaser{} + \redlaser{}.
\end{equation}

To generate a beat note between the SFG and OFC outputs, a $2\times2$ polarisation maintaining fibre coupler (Thorlabs PN635R5F2) was used to combine the SFG radiation with 1~mW OFC light amplified around 629~nm. The OFC at 629~nm was generated by second-harmonic generation (SHG) of amplified and frequency-shifted OFC radiation (Menlo Systems M-VIS extension). For improving the signal-to-noise ratio of the beat note, a holographic grating with 2400 lines/mm was used to disperse the comb spectrum in order to suppress spurious radiation impinging on the photodetector. Furthermore, a balanced photodetector (Thorlabs PDB425A-AC) was employed for rejection of common-mode noise. The total effective powers of the SFG and OFC radiation in each arm of the balanced setup after spectral overlap and filtering were measured to be 12~$\mu$W and 8~$\mu$W, respectively. Given the comb repetition rate of 250~MHz, the power of the comb mode beating with the SFG light in each arm of the setup was about 1.5~nW. 

\begin{figure}[ht] 
    \centering
    \includegraphics[width=1\linewidth,trim={0.3cm 3cm 0.6cm 3cm},clip]{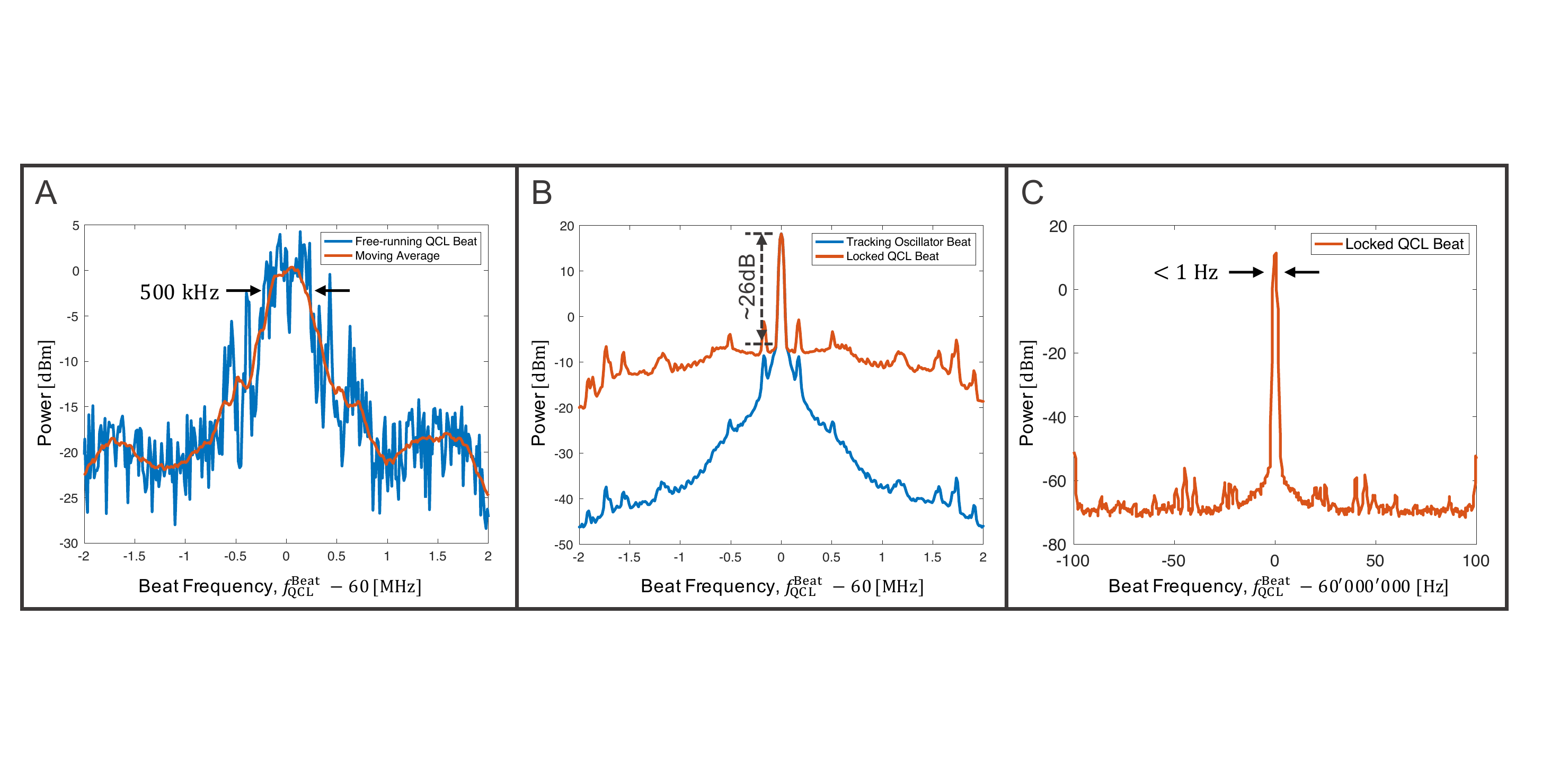}
    \caption[Locking Results]{A) Beat note of the free-running QCL with the OFC measured at a resolution bandwidth of 30~kHz (blue trace). A linewidth (FWHM ) of 500~kHz was determined for the laser from a 20-point running average of the signal (orange trace). B) In-loop beat note of the QCL locked to the OFC (orange trace). A signal-to-noise ratio of 26~dB was observed at a 30~kHz resolution bandwidth. The signal was further filtered by locking two tracking oscillators to the beat note (see text). The spectrum of one of the oscillators is shown as the blue trace. C) Beat note of the locked QCL with the OFC measured at a resolution bandwidth of 1~Hz. A FWHM linewidth of $<1$~Hz was observed, limited by the resolution of the spectrum analyser.}
    \label{fig:locking_results}
\end{figure}

Fig. \ref{fig:locking_results}A shows the beat note of the free running QCL with the OFC. Since the 729~nm laser and the OFC were locked to the ULE cavity and consequently featured much smaller linewidths than the QCL, the measured FWHM linewidth of $\sim$500~kHz was attributed to the noise of the free-running QCL with negligible contributions from fibre-noise caused by the 30~m long unstabilised fibres delivering the 729~nm and OFC light to the beat-note setup. The beat note was band-pass filtered with a bandwidth of 5~MHz. A phase-locked loop was then engaged in order to lock the QCL to the OFC. The orange trace in fig. \ref{fig:locking_results}B shows the in-loop beat note of the QCL. The observed linewidth of the locked beat note was observed to be $<1$~Hz (fig. \ref{fig:locking_results}C), limited by the resolution of the spectrum analyser, indicating that the QCL frequency closely follows the OFC and, therefore, the master laser. This implies that the linewidth of the QCL was comparable to the one of the master laser. The fibre-coupled balanced-detection scheme allowed for cancellation of common-mode noise in both arms of the detection setup and resulted in a beat note free from misalignment with an in-loop signal-to-noise ratio of 25-27~dB within a measurement bandwidth of 30~kHz. 

Because the signal-to-noise ratio of the beat note was insufficient for a direct and reliable counting of its frequency, two voltage-controlled oscillators were phase locked to the beat note. The electronic schematic of the home-built tracking oscillator can be found in ref. \cite{sinhal22a}. The bandwidths of the locks were adjusted such that the oscillators could follow the beat note without copying all of its noise. This allowed for a better filtering of the beat note. The blue trace in fig. \ref{fig:locking_results}B shows the spectrum of one of the tracking oscillators locked onto the in-loop QCL beat. The tracking oscillators were redundantly counted on two separate channels of the counter. A cycle slip was detected when the difference between the two counted frequencies exceeded a threshold of 0.5~Hz. The threshold was chosen such that all cycle slips could be reliably identified and removed without discarding results from the two counter channels that differed slightly due to measurement statistics. 

The frequency of the sum-frequency radiation could be determined from the frequencies of one of the tracking oscillators, \sfgbeat{}, as, 
\begin{equation}
    \sfglaser{} = 2\ceo{} + \sfgn{}\rep{} + \sfgbeat{},
\end{equation}
where, $\sfgn{} = 1'907'307$ was the mode number of the comb tooth beating with the sum frequency and, as before, a factor of 2 appears with \ceo{} because the spectrum of the comb was doubled for generating the beat note with the sum frequency. The frequency of the QCL could then be calculated from eq. (\ref{eqn:sfg}) as, 
\begin{equation} \label{eqn:QCLGPS}
    \qcllaser{} = 2\ceo{} + \sfgn{}\rep{} + \sfgbeat{} - \redlaser{}.
\end{equation}

Similar to the frequency determination for the master laser outlined in sec. \ref{sec:stab_ofc}, the frequency of the QCL could be accurately determined by eliminating the repetition rate of the OFC from the above equation by making use of eq. (\ref{eqn:fus}). $\qcllaser{}$ was then calculated as, 
\begin{equation} \label{eqn:QCLUS}
    \qcllaser{} = -y\ceo{} + y(\uslaser{} - \usbeat{}) + (\sfgbeat{} - \redbeat{}),
\end{equation}
where, $y = (\sfgn{} - \redn)/\usn{} \approx 0.34$.

\section{Performance of the Stabilisation Scheme}
\label{sec:perf}

\begin{figure}[ht] 
    \centering
    \includegraphics[width=0.6\linewidth,trim={1cm 7.5cm 2cm 8cm},clip]{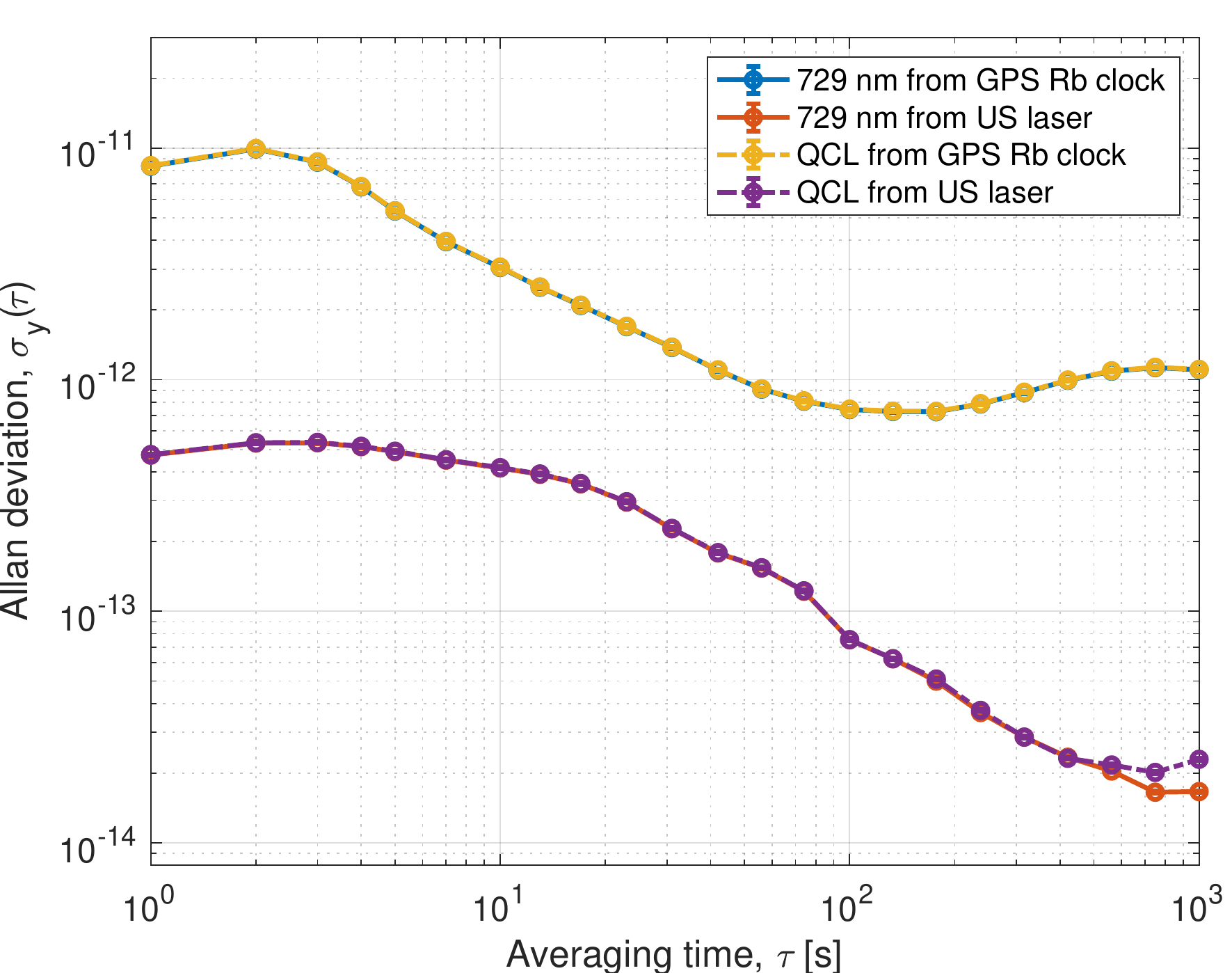}
    \caption[QCL Traceability]{Allan deviations of the 729~nm laser and the QCL frequency determined by two methods. In the blue and yellow traces, the GPSD Rb clock was employed in order to determine the frequencies of the lasers. The results of the laser frequencies when referenced to the US laser is shown in orange and violet. Slight difference between the 729~nm laser and the QCL at 800-1000~s averaging time is most likely due to measurement statistics.}
    \label{fig:qcl_traceability}
\end{figure}

Fig. \ref{fig:qcl_traceability} shows the stabilities of the lasers in terms of their Allan deviations evaluated by two methods. In the first approach, the frequencies of the 729~nm laser (blue trace) and the QCL (yellow trace) were determined by the GPSD-Rb clock (equations \ref{eqn:729GPS} and \ref{eqn:QCLGPS}, respectively). The measurements were limited by the instability of the local Rb clock and the drifts of the ULE cavity. In the second method, the frequencies of both lasers (orange and purple traces) were determined using eqs. (\ref{eqn:729US}) and (\ref{eqn:QCLUS}) by referencing to the remote US laser. Here, the repetition rate of the OFC was eliminated and the limiting factor was the instability of the US laser frequency. Additionally, the slow feedback for the compensation of the OFC was engaged resulting in the elimination of the ULE cavity drifts and in a further averaging down of the instability of both lasers. 

The reference to the US laser enabled a reduction of the instability of the lasers of more than one order of magnitude over the averaging time adopted in fig. \ref{fig:qcl_traceability}. We measured a frequency instability of $5\times10^{-13}$ at 1~s and $2\times10^{-14}$ at 1000~s integration time for both lasers. The Allan deviations of the QCL replicate that of the 729~nm in both methods indicating a tight lock of the QCL to the OFC. We do not observe systematic contributions to the QCL frequency due to uncompensated noise from fibres delivering the 729~nm and the OFC light to the QCL stabilisation setup or due to the additional tracking oscillators at these levels. 

\section{Discussion}
\label{sec:disc}

As demonstrated in fig. \ref{fig:qcl_traceability}, the long-term stability of the QCL frequency closely follows the one of the 729~nm master laser. The latter is determined by the instability of the ultrastable frequency reference transferred via the fibre link and has been demonstrated to be as small as $\approx 10^{-15}$ at 1000~s integration time \cite{husmann21a}. Note that these stabilities reported in fig. \ref{fig:qcl_traceability} are about one order of magnitude smaller than that figure due to a technical limitation at METAS at the time of the measurements. 

Similarly, from the linewidths of the beat notes shown in figs. \ref{fig:729lock}B and \ref{fig:locking_results}C, we can surmise that the short-term linewidth of the QCL output closely mimics the one of the 729~nm master laser. Due to lack of an even more precise linewidth reference in our laboratory, we can currently only give an upper bound of $<2$~kHz FWHM for the linewidth of the master laser derived from the data shown in fig. \ref{fig:USlock} which was limited by fibre noise originating from the frequency transfer of the US laser. However, because the short-term linewidth of the master laser is governed by the PDH lock to the ULE cavity, we expect it, and consequently the linewidth of the QCL, to be considerably narrower than 2~kHz in the current setup. A stringent test of the QCL linewidth can be obtained by, e.g., spectroscopic measurements of ultranarrow rovibrational transitions in N$_2^+$ which are being undertaken in our laboratory \cite{germann14a}.

The present scheme can be compared with previously reported approaches for QCL stabilisation. In the work of Hansen et al. \cite{hansen13a}, upconverted QCL radiation was referenced to an OFC. A linewidth reduction was not implemented in that study so that the absolute QCL frequency could be determined at the 100~kHz level with a frequency stability of a few kHz. An analogy can also be drawn to the setup by the same authors reported in ref. \cite{hansen15a} in which the upconverted QCL radiation was narrowed by a second ULE cavity. Our scheme, on the other hand, directly utilises the upconverted radiation for both linewidth narrowing and absolute referencing to an SI traceable OFC thus reducing the experimental complexity. 

The present setup can also be contrasted with the one of Argence et al. \cite{argence15a} in which a shifted OFC was generated by nonlinear mixing of the OFC light with radiation from QCL. This shifted comb was then used for the phase locking and absolute referencing of the QCL by comparison to a remote SI-traceable US reference. Here, we rely on the transfer of high spectral purity of a 729~nm laser in the local laboratory locked to a high-finesse optical cavity to a QCL via an OFC. By utilising a local laser for the short-term stabilisation, we ensure that the apparent linewidth of US lasers generated in remote locations and transferred to local laboratory via long phase-stabilised fibre links with limited fibre-noise cancellation bandwidths do not affect the linewidth of the lasers in the local laboratories. In our setup, the SI-traceable US laser was used only for the long-term stabilisation and absolute-frequency stabilisation of the local lasers.

\section{Conclusion and outlook}
\label{sec:conc}

We have presented an approach for the frequency stabilisation, absolute frequency determination and linewidth narrowing of a MIR QCL. The high spectral purity of a NIR laser locked to a high-finesse ULE cavity was transferred to the QCL via an OFC following frequency upconversion of the MIR radiation. Long-term stabilisation and absolute frequency determination of the QCL was achieved by comparison to a SI-traceable US reference generated in a remote location and disseminated via a phase-stabilised fibre link. The techniques for implementing tracking oscillators and redundant counting employed here are robust and can readily be incorporated in conventional beat-note detection chains. They facilitate the accurate counting of beat-note frequencies even in situations in which the signal-to-noise ratios of the beat notes are otherwise insufficient. Frequency-stabilised QCLs such as the one presented here are essential components for the precise spectroscopy of molecules and for the implementation of molecular clocks and qubits in which the linewidth and long-term stability the laser frequency often limits the achievable accuracy and precision. The presented approach is flexible and can be extended to QCLs operating in the entire MIR spectral domain by a proper choice of the non-linear process for frequency conversion.  

\section{Acknowledgements}

We thank Drs. Dominik Husmann and Jacques Morel for providing the US reference frequency from METAS and for helpful discussions regarding the stabilisation setup. Further discussions are  acknowledged with Dr. Cecilia Clivati (INRIM), Dr. Maximilian Bradler (Menlo Systems), Prof. J\'er\^ome Faist (ETH Zurich) and Prof. Fr\'ed\'eric Merkt (ETH Zurich). We thank Georg Holderied for technical support with the electronics. We acknowledge funding from the Swiss National Science Foundation as part of the National Centre of Competence in Research, Quantum Science and Technology (NCCR-QSIT) as well as grant nrs. CRSII5\_183579 and 200021\_204123.

\section{Data availability}

The data that support the findings of this study have been deposited in Zenodo with the identifier DOI: https://doi.org/10.5281/zenodo.7035071.


\end{document}